\documentclass{JINST}
\usepackage{mathtools}
\usepackage[]{placeins}
\usepackage[]{subcaption}
\usepackage{multirow}
\usepackage[]{threeparttable}

\title{The $\mu$TPC Method: Improving the Position Resolution of Neutron Detectors Based on MPGDs }

\author{
Dorothea~Pfeiffer$^{a,b}$\thanks{Corresponding
author.}$\ $,
Filippo~Resnati$^{a,b}$\thanks{Corresponding
author.}$\ $,
Jens~Birch$^{c}$,
Richard~Hall-Wilton$^{a,d}$, 
Carina~H\"{o}glund$^{a,c}$,
Lars~Hultman$^{d}$,
George~Iakovidis$^{b,e}$,
Eraldo~Oliveri$^{b}$,
Esko~Oksanen$^{a}$,
Leszek~Ropelewski$^{b}$,
Patrik~Thuiner$^{b,f}$
\\
\llap{$^a$}European Spallation Source (ESS AB)\\
	P.O. Box 176, SE-22100 Lund, Sweden\\
\llap{$^b$}CERN\\
  	CH-1211 Geneva 23, Switzerland\\
\llap{$^c$}Link\"{o}ping University\\
	IFM SE-581 83 Link\"{o}ping, Sweden\\
\llap{$^d$}Mid-Sweden University\\
	SE-85170 Sundsvall, Sweden\\
\llap{$^e$}Brookhaven National Laboratory\\
	P.O. Box 5000, Upton, 11973-5000, USA\\
\llap{$^f$}Vienna University of Technology\\
	1040 Vienna, Austria\\

E-mail: \email{Dorothea.Pfeiffer@cern.ch, Filippo.Resnati@cern.ch}}

\abstract{Due to the $^3$He crisis, alternatives to the \emph{standard} neutron detection techniques are becoming urgent.
In addition, the instruments of the European Spallation Source (ESS) require advances in the state of the art of neutron detection.
The instruments need detectors with excellent neutron detection efficiency, high rate capabilities and unprecedented spatial resolution.
The Macromolecular Crystallography instrument (NMX) requires a position resolution in the order of 200 $\mu$m over a wide angular range of incoming neutrons. Solid converters in combination with Micro Pattern Gaseous Detectors (MPGDs) are proposed to meet the new requirements. Charged particles rising from the neutron capture have usually ranges larger than several millimetres in gas. This is apparently in contrast with the requirements for the position resolution. In this paper, we present an analysis technique, new in the field of neutron detection, based on the Time Projection Chamber (TPC) concept. Using a standard Single-GEM with the cathode coated with $^{10}$B$_4$C, we extract the neutron interaction point with a resolution of better than $\sigma = 200$~$\mu$m.
}

\begin{document}
\section{Introduction}\label{sec:intro}
The European Spallation Source (ESS)~\cite{ESS} in Lund/Sweden is foreseen to be operational in 2019. It will become the world's most powerful thermal neutron source, with a significantly higher brightness than existing reactor sources like the Institut Laue-Langevin (ILL)~\cite{ILL} or other spallation sources like the Spallation Neutron Source (SNS)~\cite{SNS} and the Japan Proton Accelerator Research Complex (J-PARC)~\cite{J-PARC}. Currently 22 neutron scattering instruments are planned as the baseline suite for the facility~\cite{ESS_TechnicalDesignReport}. Efficient thermal neutron detectors are a crucial component for each of the instruments~\cite{Vertex}. The dominant detector technology at neutron scattering instruments so far has been gaseous $^3$He detectors~\cite{ILL_Neutron}. Due to the $^3$He crisis~\cite{He3_crisis1, He3_crisis2} however, an extensive international R\&D program is currently under way in order to develop efficient and cost-effective detectors based on other isotopes~\cite{ICND,Zeitelhack, Guerard}. The solid isotope $^{10}$B has a significant neutron capture cross section and might thus be an alternative to $^3$He as neutron converter. The charged particles created in a solid neutron converter have subsequently to be extracted and electrically detected. For large area detectors, the only realistic choices are gaseous detectors like gas proportional counters surrounded by thin films of $^{10}$B-enriched boron carbide~\cite{B4cfilms,B10_multigrid, Bigault}. 

Some ESS instruments like the Macromolecular Diffractometer (NMX)~\cite{ESS_TechnicalDesignReport} though do not necessarily require a large detector area. Whereas macromolecular crystallography instruments at reactor sources typically use neutron image plates~\cite{ImagePlate} with ca 200 $\mu$m spatial resolution, spallation source instruments require time resolution that the image plates lack altogether. Scintillation based detectors~\cite{IBIX, MANDI} are currently limited to ca 1 mm spatial resolution. For these instruments solid converters in combination with Micro Pattern Gaseous Detectors (MPGDs)~\cite{MPGD} are a promising option to achieve the spatial resolution of the neutron image plate with time resolution, high-rate capabilities and a good neutron detection efficiency. The position resolution of a MPGD with a solid neutron converter depends on three factors: The average range and the direction of the charged particle in the gas, the diffusion of the charge and the pitch of the read-out strips. In the case of a $^{10}$B layer, a neutron is captured by the $^{10}$B nucleus, resulting in an $\alpha$ particle and a $^7$Li ion that are emitted back to back:
\begin{align*}
^{10}B + n &\rightarrow ^{7}Li^{*} +\alpha + \gamma(0.48~MeV),~Q=2.3~MeV~(93\%)\\
^{10}B + n &\rightarrow ^{7}Li~+\alpha,~Q=2.79~MeV~(7\%)
\end{align*}
The $\alpha$ particle has an energy of E$_{kin}$ \textless 1.78 MeV when it enters the active gas volume. At this kinetic energy the average range amounts to about 8 mm in Ar/CO$_{2}$ (70$\%$/30$\%$) at atmospheric pressure. A reconstruction of the position of the impinging neutron based on the centroid of the track results in an error of 50$\%$ of the projected track length. The $\mu$TPC concept has been developed by the ATLAS collaboration and is used in particle physics~\cite{uTPC, uTPC2}. Its application leads to a clear improvement in the reconstruction of the positions of muons detected with Micromegas. A position resolution of 100 $\mu$m independent of the angle of incidence is reached. By applying this type of analysis for the first time to neutron detectors, it is possible to reliably identify the beginning of all $\alpha$ and $^{7}$Li ion tracks and thus to reconstruct the position of the incident neutron at the required resolution for the NMX instrument. This work consists of a description of the setup, the measurement and the data analysis that was performed in order to retrieve the full information on the position of the neutron conversion.

\section{Experimental setups}
\label{sec:setup}
The detectors used for the measurements were one standard 10 x 10~cm Triple-GEM~\cite{GEM} and one 10 x 10~cm Single-GEM detector. 
\subsection{Triple-GEM detector setup}
\begin{figure}[htbp]
\centering
\includegraphics[width=0.8\textwidth]{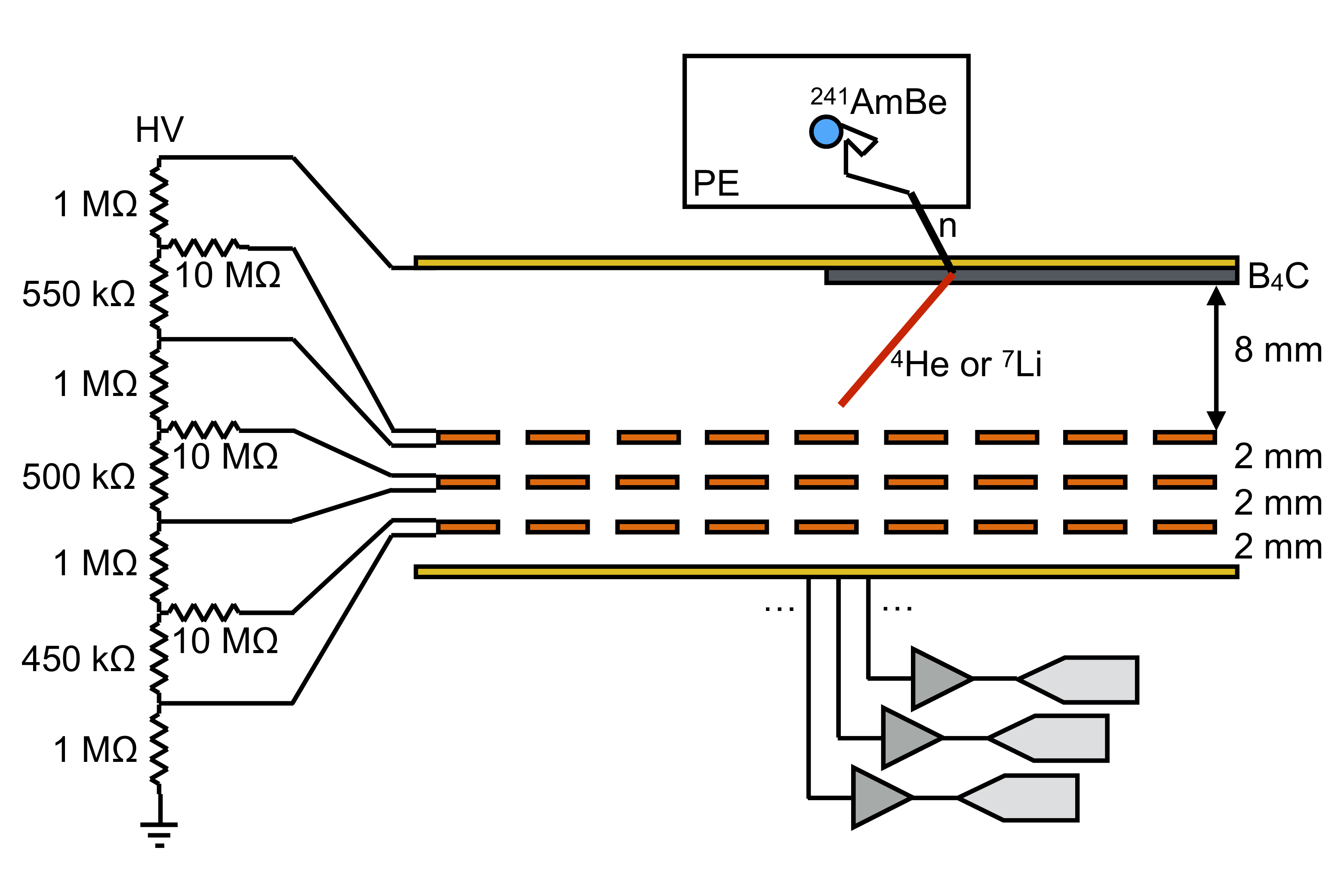}
\caption{Schematic representation of the setup for a Triple-GEM detector configuration.}
\centering
\label{fig: measurement_setup}
\end{figure}

The Triple-GEM setup is sketched in figure~\ref{fig: measurement_setup}.
A solid layer of $^{10}$B$_4$C of 1~$\mu$m thickness sputtered onto a 25~$\mu$m thick aluminium foil~\cite{B4cfilms, radhard} was used as neutron converter. It covered only half of the active area of the detector, and was kept at the same potential as the standard detector cathode installed above. The drift length of the detector was 8~mm, long enough to fully contain the $\alpha$ particles and $^7$Li ions from the $^{10}$B(n, $\alpha$)$^7$Li and $^{10}$B(n, $\alpha \gamma$)$^7$Li reactions. At room temperature and atmospheric pressure, the entire detector volume was constantly flushed with a flow of 3~l/h with Ar/CO$_2$~70/30 mixture finally released into the atmosphere. The high voltage was supplied via a resistor divider (see values in figure~\ref{fig: measurement_setup}). The current through the divider was set to 565 $\mu$A in order to obtain an effective gain of about 60. A low gain is sufficient due to the large amount of energy deposited by the ions of interest in the gas. The drift field of 700 V/cm was chosen to avoid electron attachment to electronegative impurities and the subsequent loss of primary ionization electrons, while keeping the drift velocity smaller than 2.0~cm/$\mu$s~\cite{Drift}. 

\subsection{Single-GEM detector setup}
A 300~$\mu$m thick aluminium sheet coated with 1~$\mu$m of $^{10}$B$_4$C ~\cite{B4cfilms} was used as neutron converter and detector cathode of the Single-GEM. As shown in figure~\ref{fig: cathode}, copper tape of 50~$\mu$m thickness was attached to the $^{10}$B$_4$C layer to stop the $\alpha$ particles and $^7$Li ions from leaving the $^{10}$B$_4$C and entering the gas, thus providing a sharp edge to determine the position resolution.  For the measurements with the Single-GEM a gain of 40 was used in combination with a drift gap of 8 mm and a drift field of 1 kV/cm. At this drift field the drift speed of electrons amounts to 3.5~cm/$\mu$s in Ar/CO$_{2}$ (70$\%$/30$\%$)~\cite{Drift}. Sufficiently low drift velocities assure the ability to distinguish the inclination of the $\alpha$ particle and $^7$Li ion tracks, allowing their three-dimensional reconstruction. 

\begin{figure}[htbp]
\centering
\begin{subfigure}[b]{.49\textwidth}
\centering
\includegraphics[width=\linewidth]{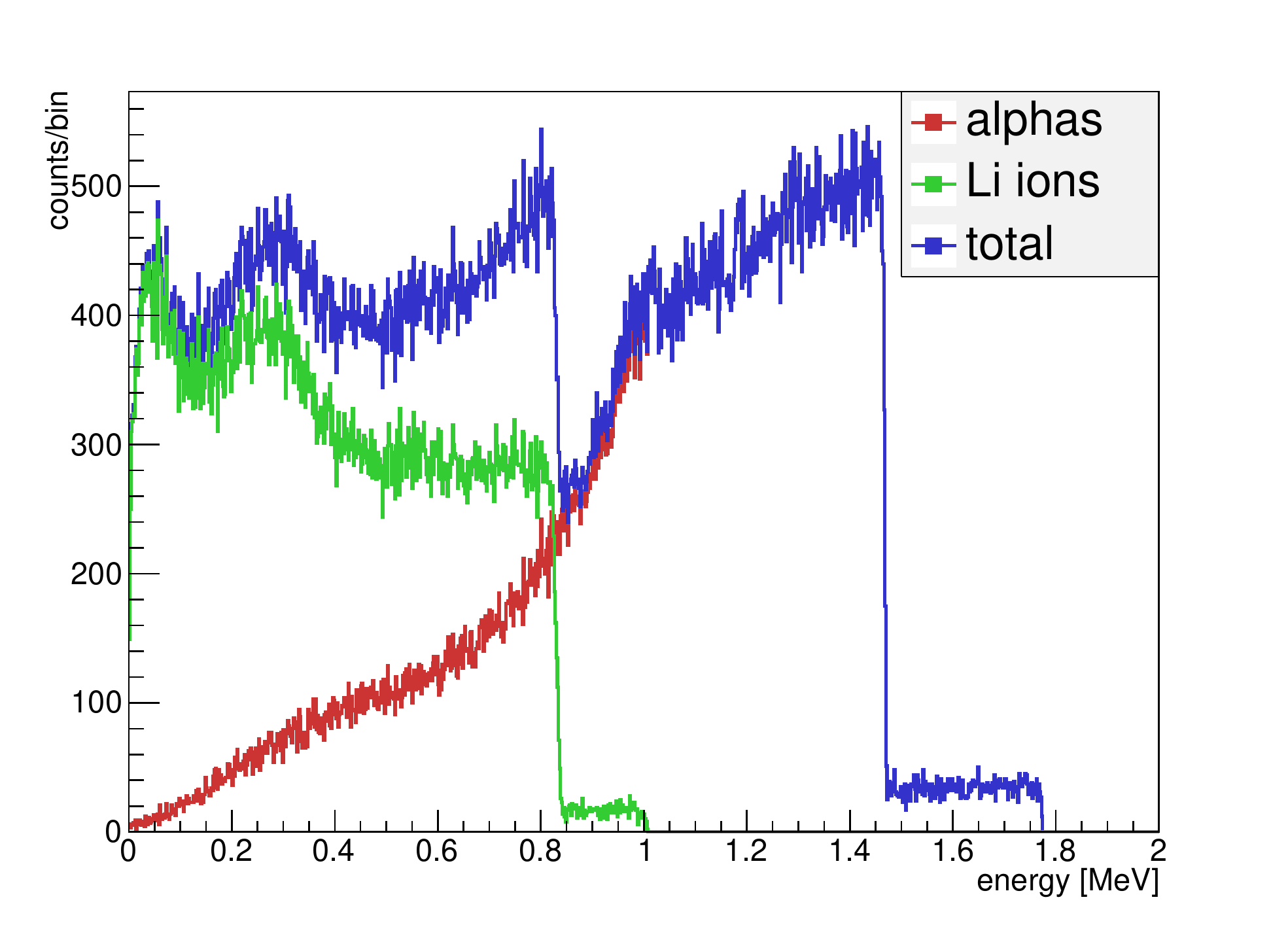}%
\caption{Spectrum simulated with Geant4}
\label{fig: spectrum_simulated}
\end{subfigure}%
\begin{subfigure}[b]{.51\textwidth}
\centering
\includegraphics[width=\linewidth]{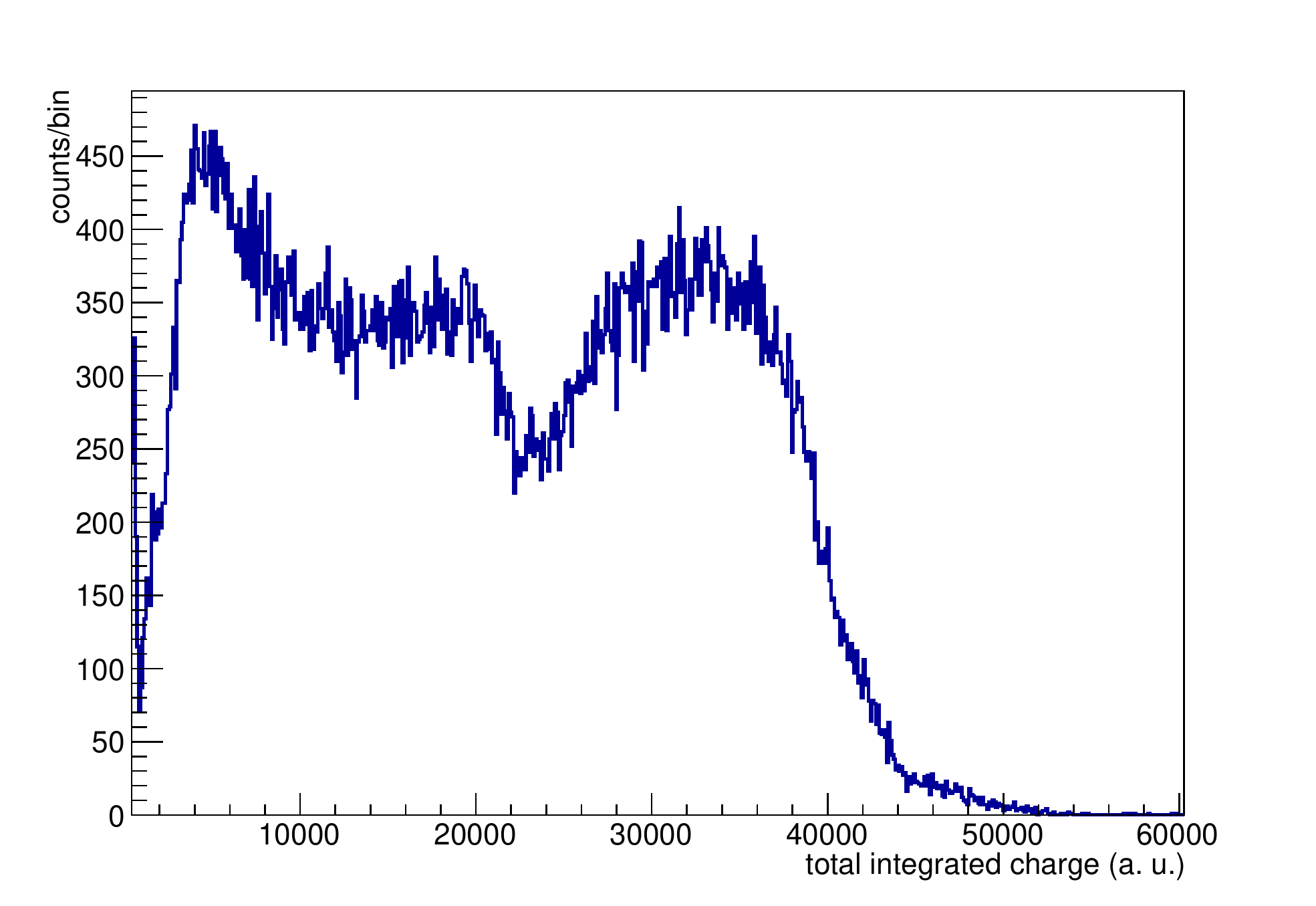}%
\caption{Spectrum measured with Triple-GEM}
\label{fig: spectrum_measured}
\end{subfigure}
\caption{Energy deposited in the drift volume by the charged particles from the $^{10}$B$_4$C neutron capture.}
\label{fig: spectra}
\end{figure}

\subsection{Detector readout and DAQ}
The readout board was a cartesian x/y strip readout \cite{GEM_rate} with 256 strips and 400~$\mu$m pitch in x and y direction in both setups. Four APV-25~\cite{APV} hybrid chips per readout were used to pre-amplify the signals. The waveforms were digitized with the Scalable Readout System (SRS)~\cite{SRS} and acquired with the ALICE DAQ system DATE~\cite{DATE} and the Automatic MOnitoRing Environment (AMORE)~\cite{AMORE} software. Thermal neutrons were obtained by moderating fast neutrons from a 370~MBq $^{241}$AmBe source with a 5~cm thick polyethylene block. GEANT4~\cite{Geant4a,Geant4_ESS} simulations show that the $^{241}$AmBe source produces isotropically about $2.3\times10^4$~n/s. After the moderation, about 100 thermal neutrons per second reach the 10 x 10~cm active area of the detector. Given the low conversion efficiency of the 1~$\mu$m thick B$_4$C layer, the detection rate was less than 5 Hz. Figure~\ref{fig: spectra} shows the simulated and measured spectra of the energy released in the gas volume by the charged particles created during the $^{10}$B$_4$C neutron capture. Within the energy resolution of the Triple-GEM detector, the simulated spectrum could be reproduced.

\section{TPC analysis}
\label{sec:analysis}
\begin{figure}[htbp]
\centering
\begin{subfigure}[b]{.33\textwidth}
\centering
\includegraphics[width=\linewidth]{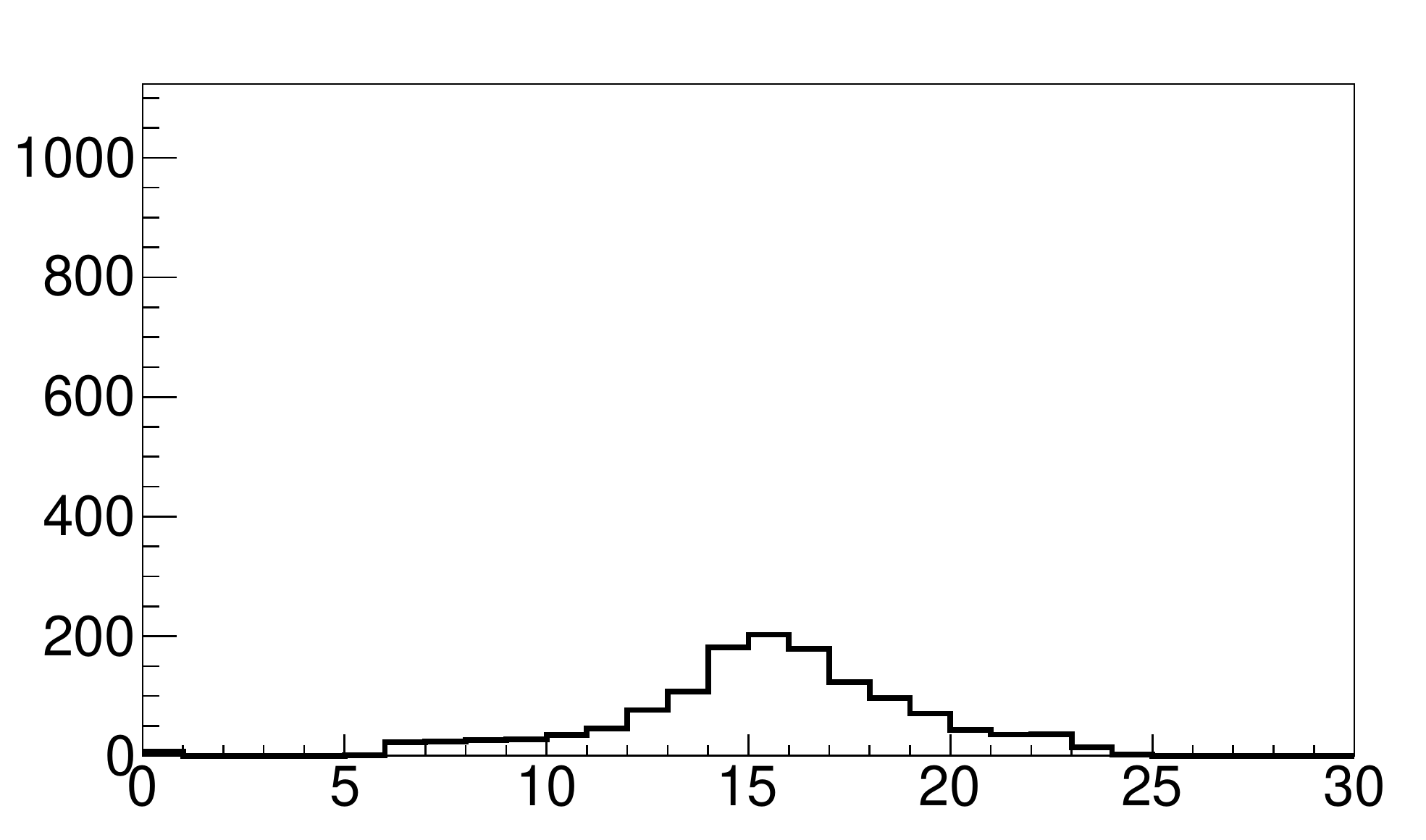}%
\caption{strip 1}
\label{fig: time_structure_x1}
\end{subfigure}%
\begin{subfigure}[b]{.33\textwidth}
\centering
\includegraphics[width=\linewidth]{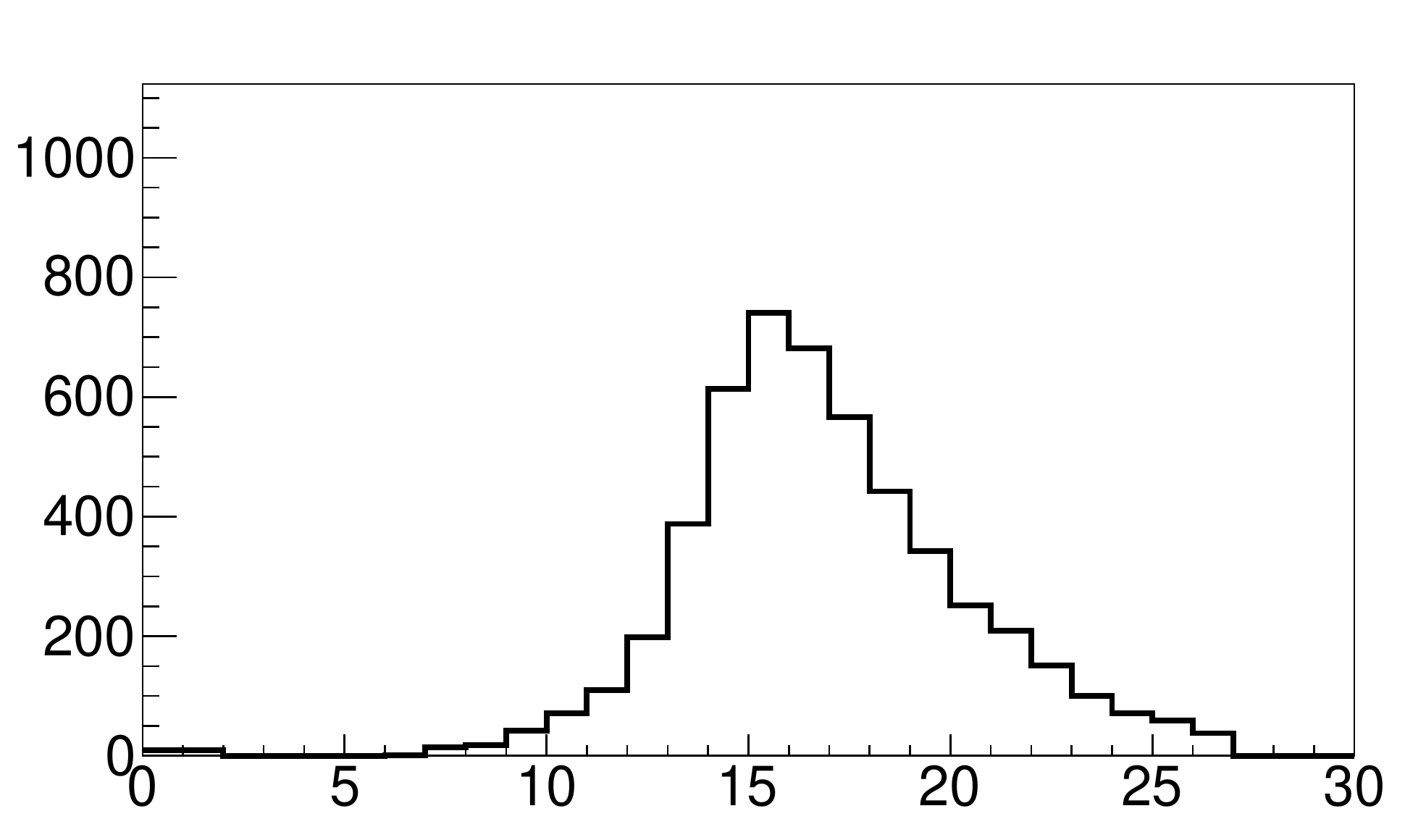}%
\caption{strip 2}
\label{fig: time_structure_x2}
\end{subfigure}%
\begin{subfigure}[b]{.33\textwidth}
\centering
\includegraphics[width=\linewidth]{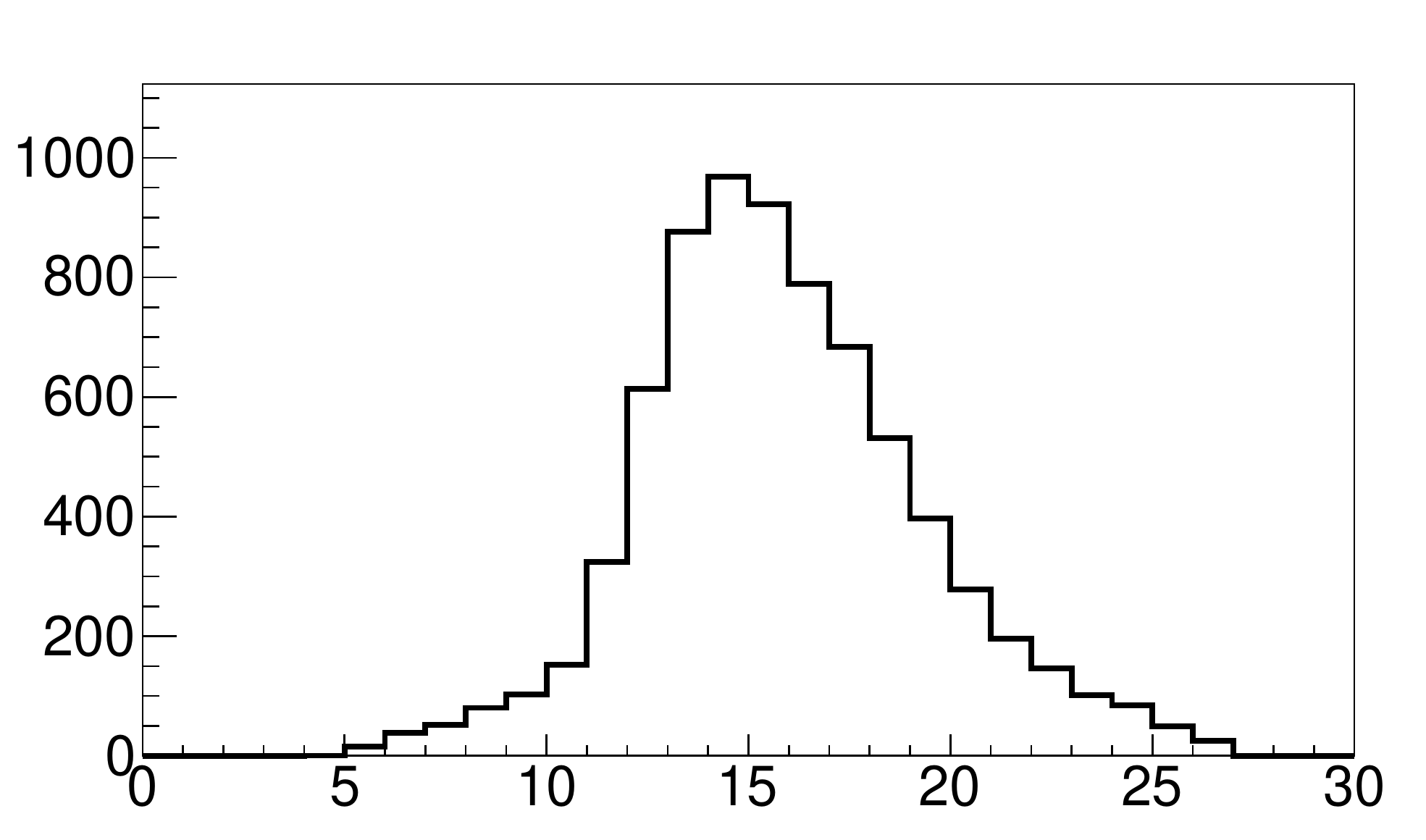}%
\caption{strip 3}
\label{fig: time_structure_x3}
\end{subfigure}%
\\
\begin{subfigure}[b]{.33\textwidth}
\centering
\includegraphics[width=\linewidth]{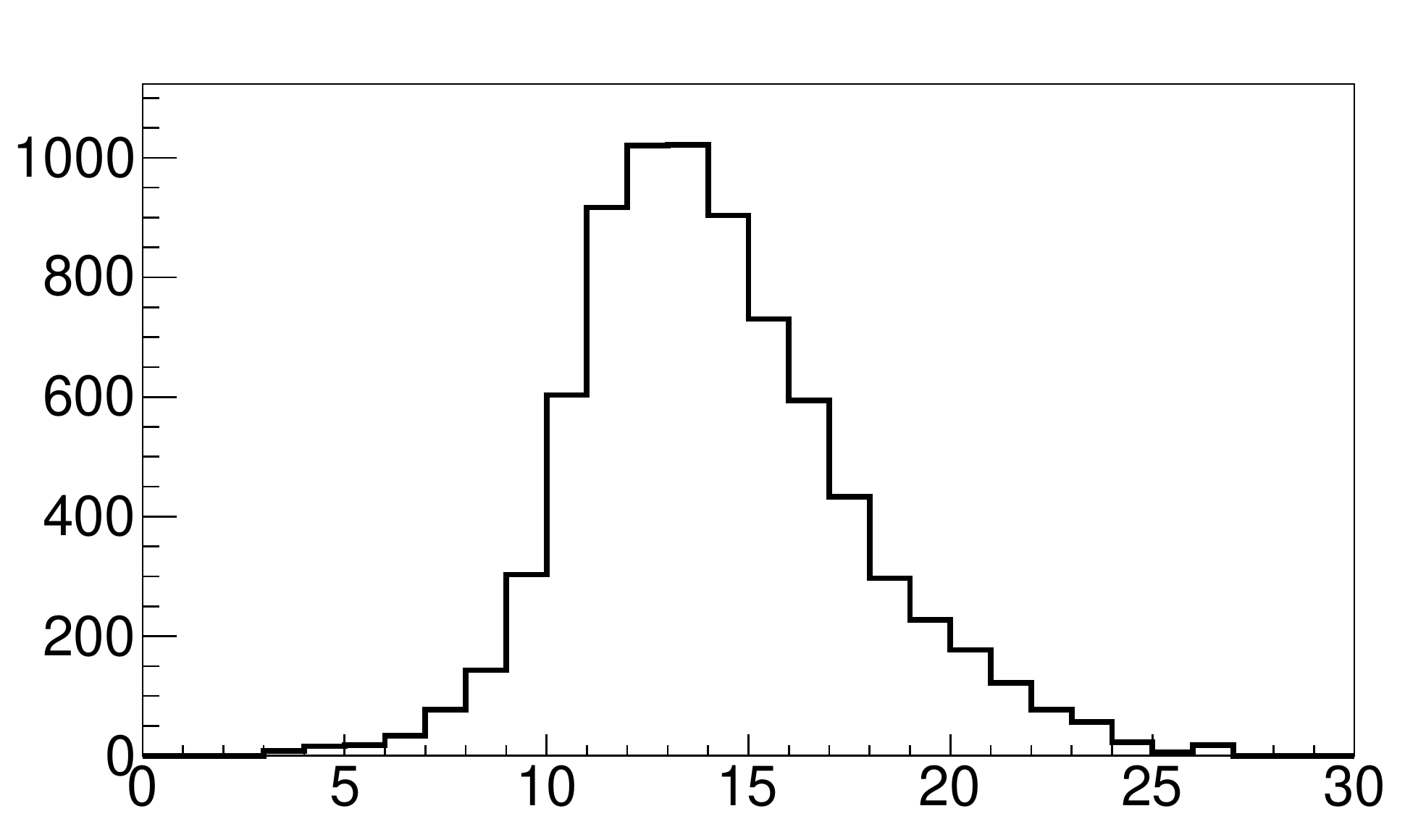}%
\caption{strip 4}
\label{fig: time_structure_x4}
\end{subfigure}%
\begin{subfigure}[b]{.33\textwidth}
\centering
\includegraphics[width=\linewidth]{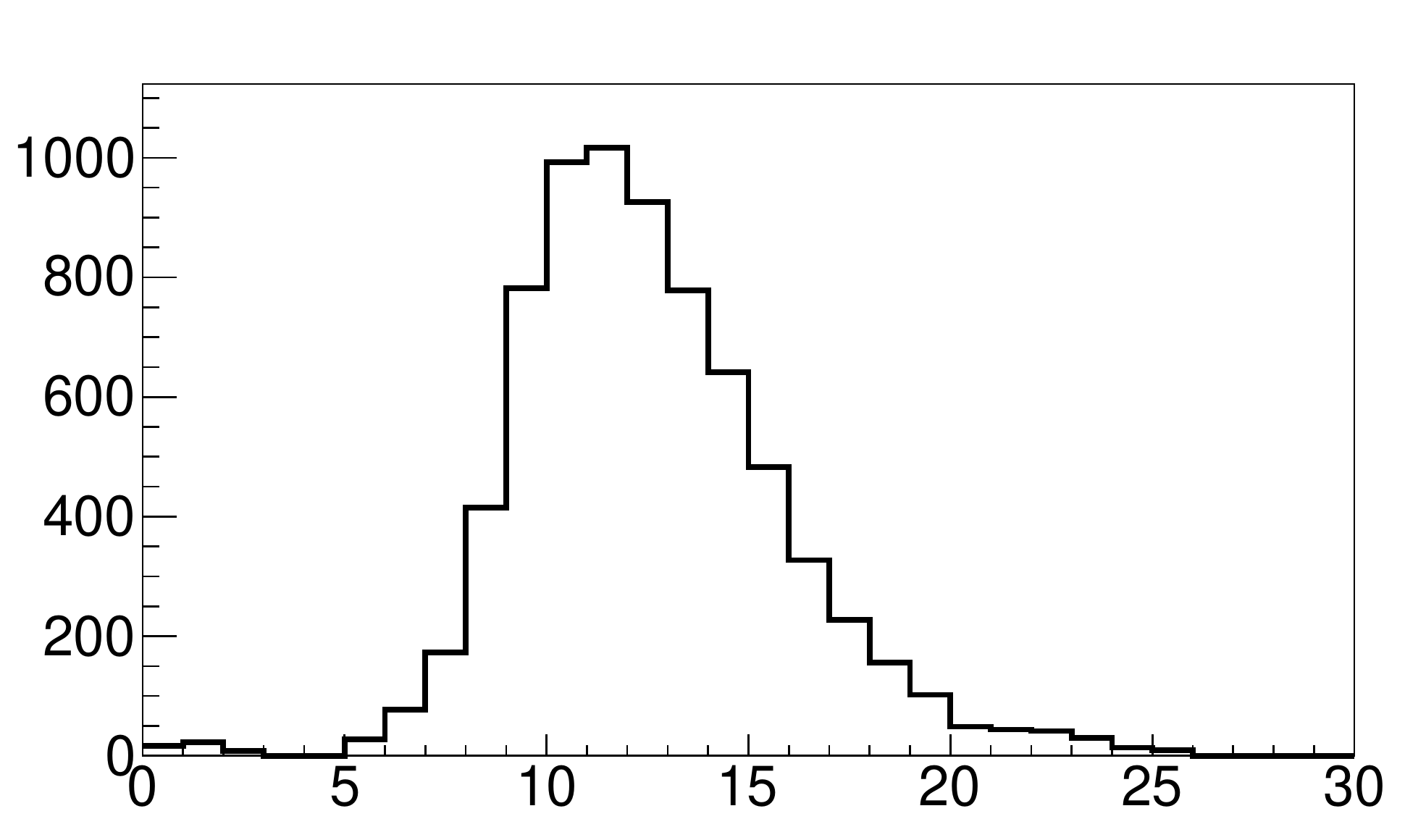}%
\caption{strip 5}
\label{fig: time_structure_x5}
\end{subfigure}%
\begin{subfigure}[b]{.33\textwidth}
\centering
\includegraphics[width=\linewidth]{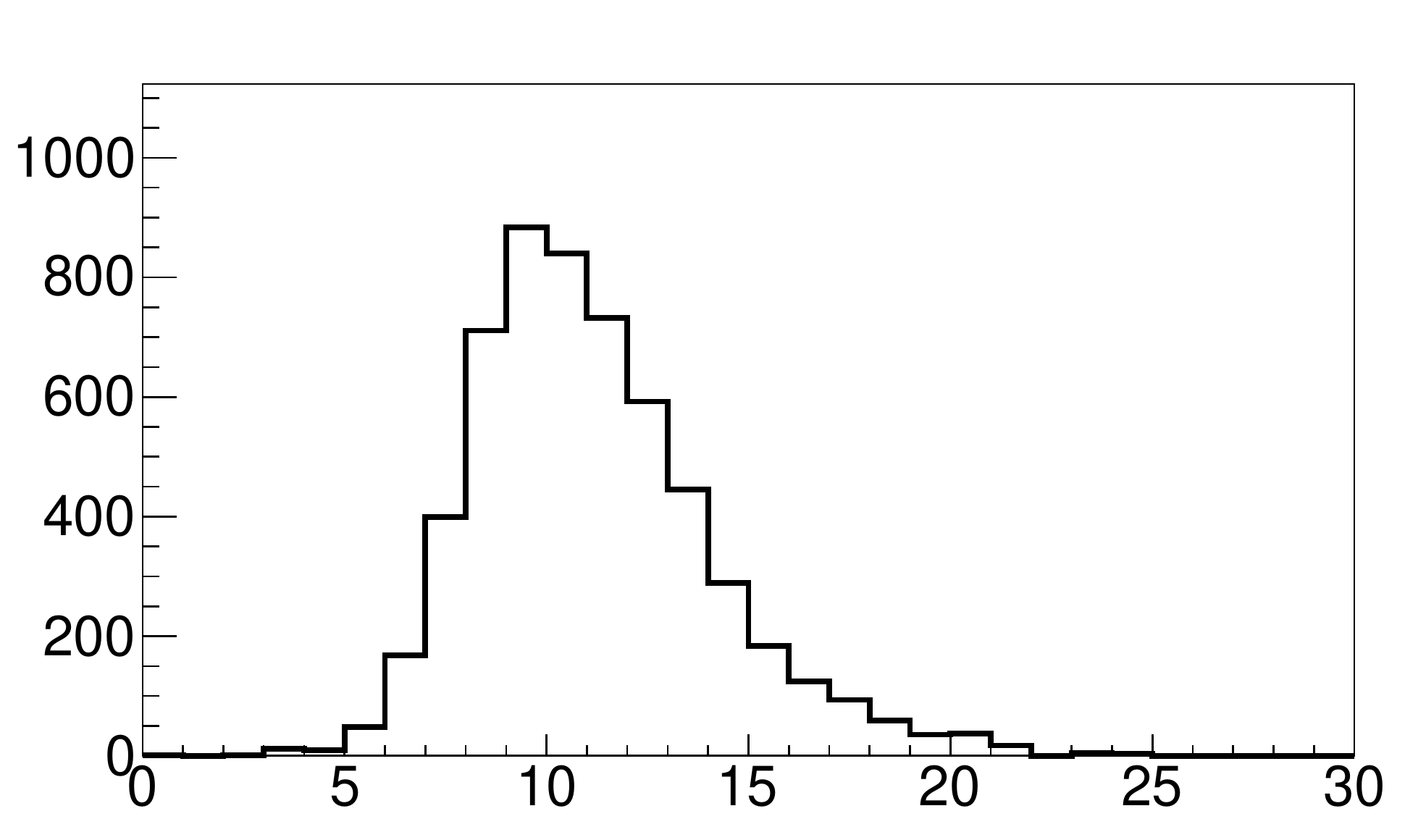}%
\caption{strip 6}
\label{fig: time_structure_x6}
\end{subfigure}%
\\
\begin{subfigure}[b]{.33\textwidth}
\centering
\includegraphics[width=\linewidth]{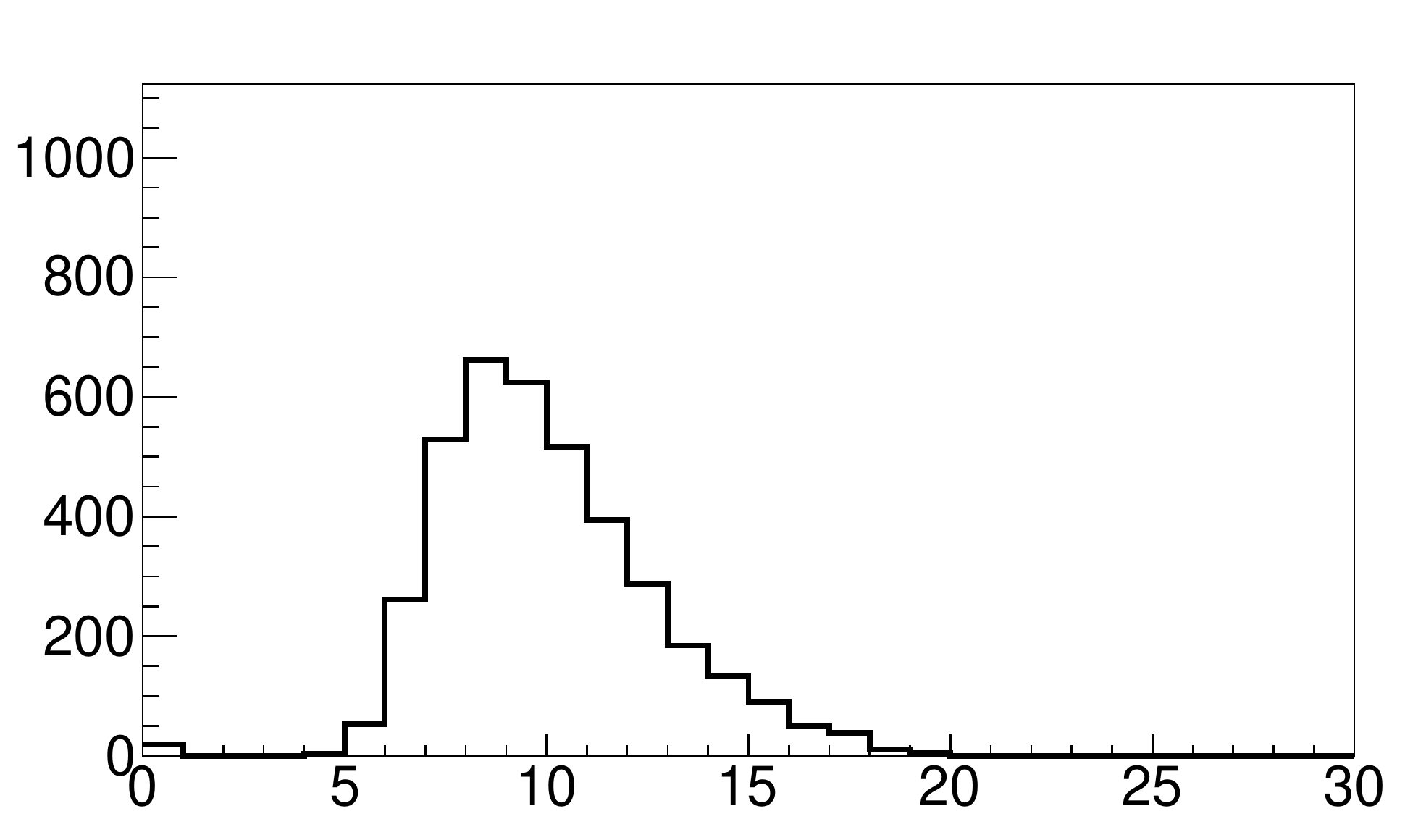}%
\caption{strip 7}
\label{fig: time_structure_x7}
\end{subfigure}%
\begin{subfigure}[b]{.33\textwidth}
\centering
\includegraphics[width=\linewidth]{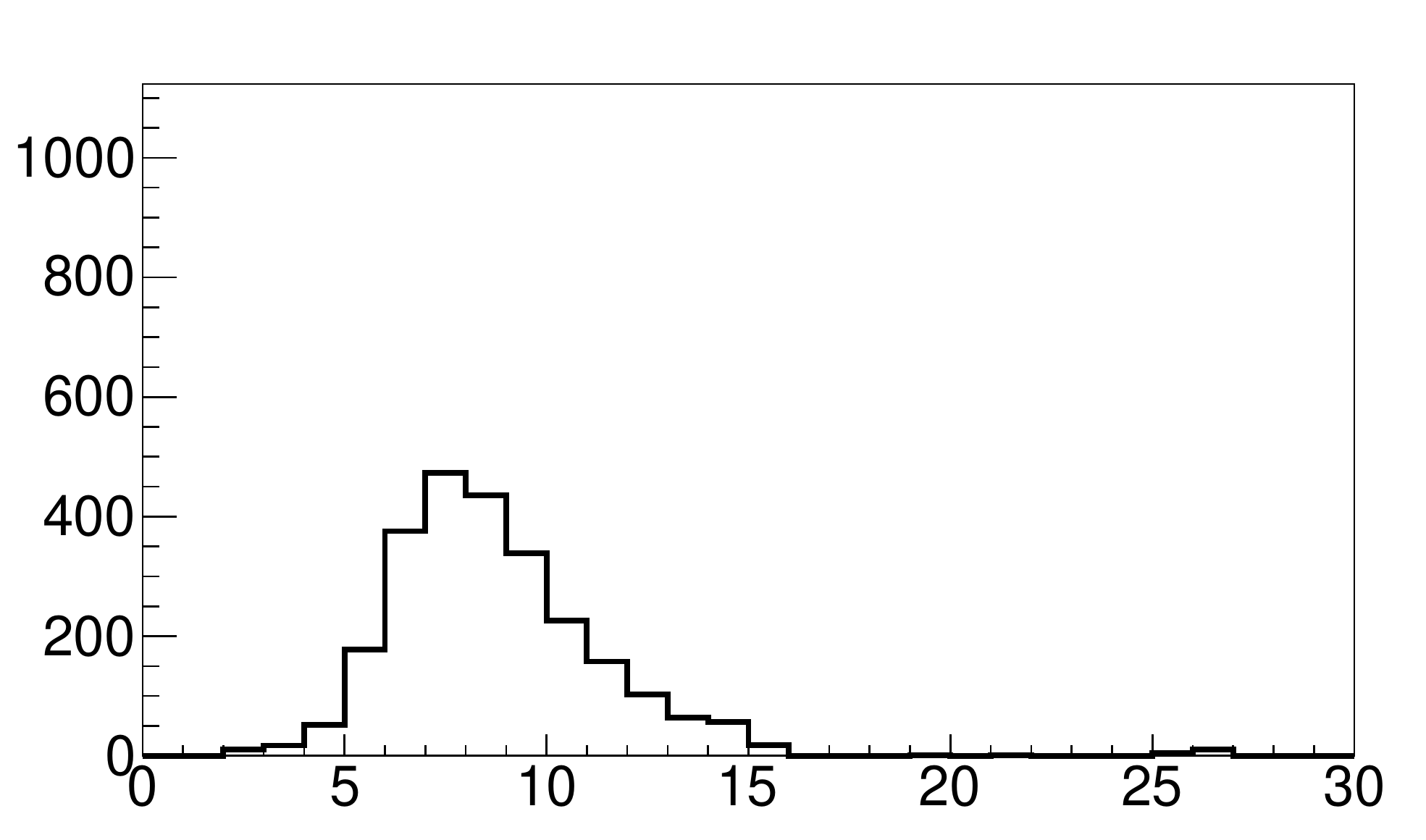}%
\caption{strip 8}
\label{fig: time_structure_x8}
\end{subfigure}%
\begin{subfigure}[b]{.33\textwidth}
\centering
\includegraphics[width=\linewidth]{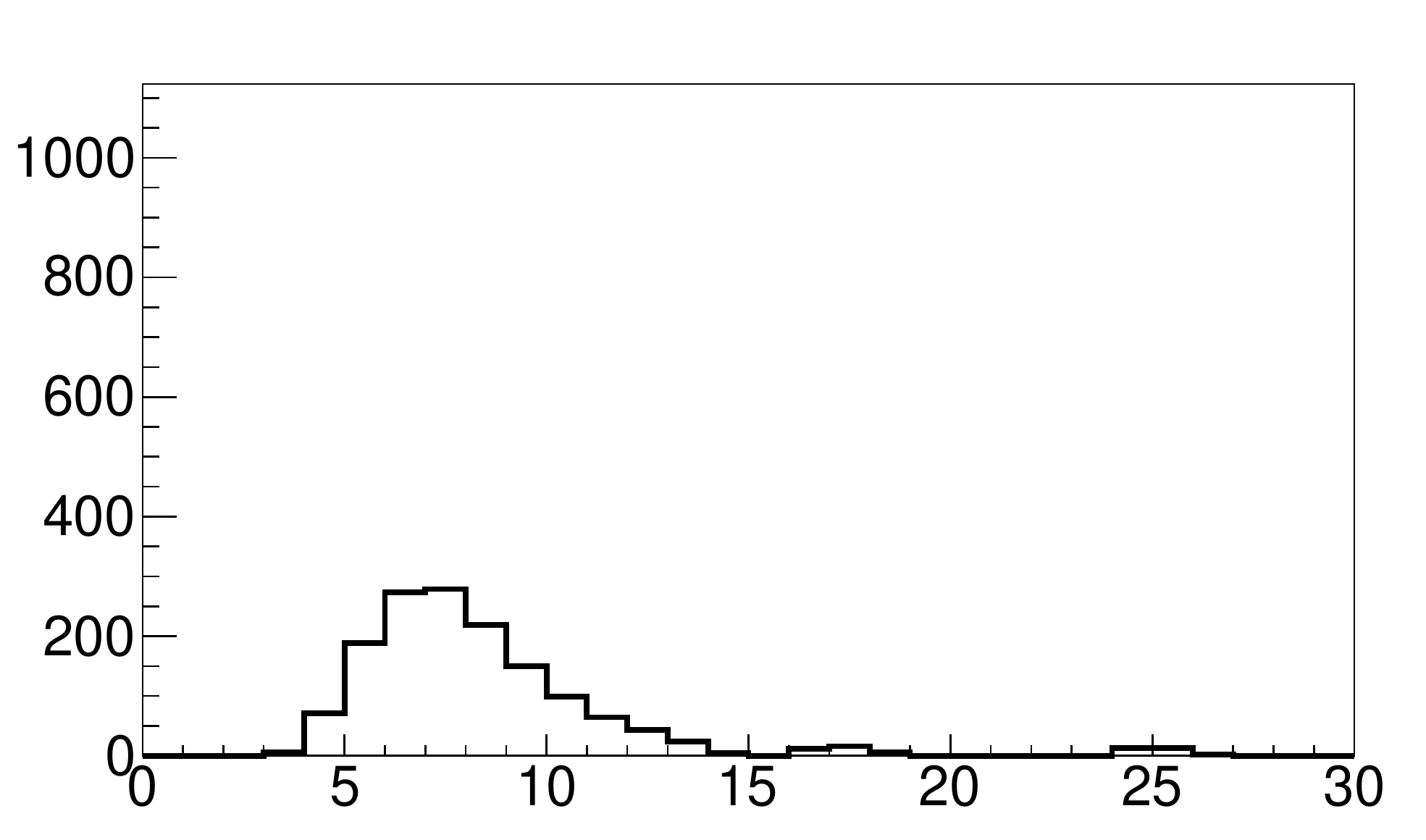}%
\caption{strip 9}
\label{fig: time_structure_x9}
\end{subfigure}%
\caption{Typical time structure of the neutron conversion signal in different strips with the 25 ns time bins on the x-axis, and the charge in arbitrary units on the y-axis.}
\label{fig: time_structure}
\end{figure}

The typical signals due to a neutron conversion are shown in figure~\ref{fig: time_structure}. The figure shows the raw waveforms from the x view on different strips, digitized every 25~ns as measured with the Triple-GEM setup. The signals on different strips are not synchronous because different drift depths require different drift times.

During the drift in electric fields, electrons diffuse as a result of multiple collisions with the gas molecules, and the initially localized charge cloud spreads out. The diffusion depends on the gas mixture, the pressure and the electrical field. At normal temperature and pressure (NTP), the $\sigma$ of the transversal and longitudinal diffusion for both detectors amounted to around 130~$\mu$m over the 8 mm drift~\cite{Gas_detectors}. 

\begin{figure}
\centering
\begin{subfigure}[b]{.5\textwidth}
\centering
\includegraphics[width=\linewidth]{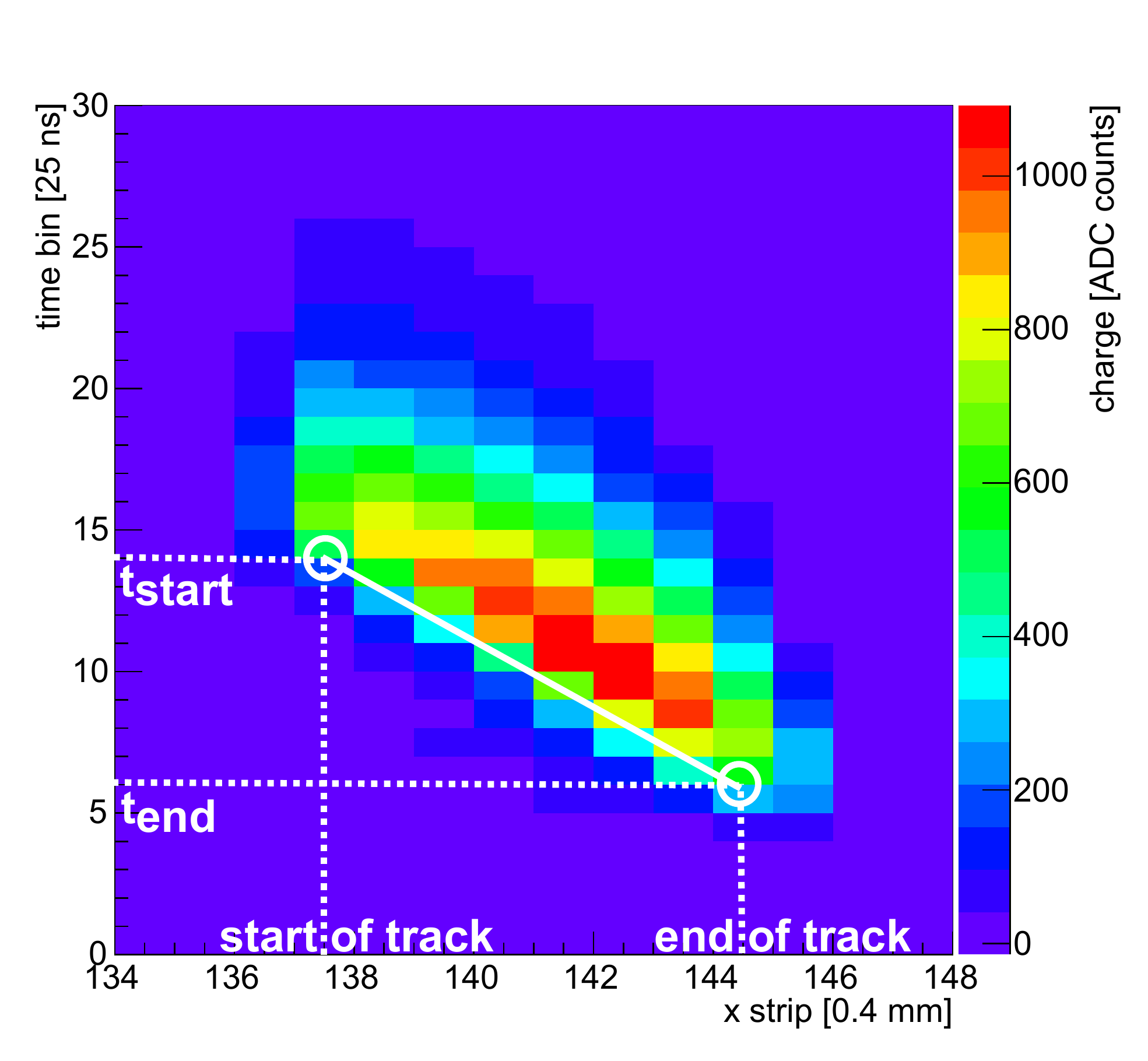}%
\caption{x view}
\label{fig: track_x_3gem}
\end{subfigure}%
\begin{subfigure}[b]{.5\textwidth}
\centering
\includegraphics[width=\linewidth]{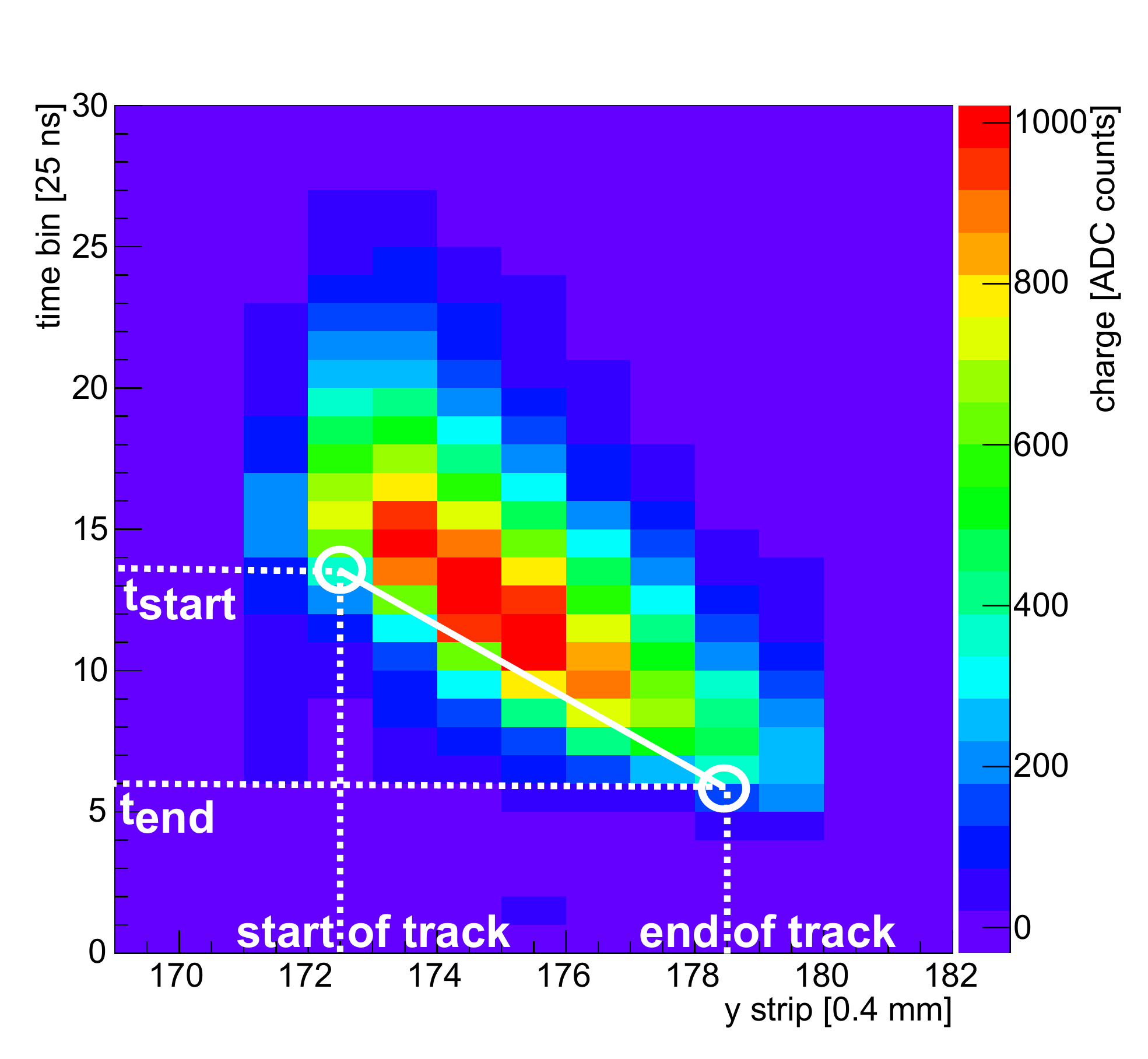}%
\caption{y view}
\label{fig: track_y_3gem}
\end{subfigure}
\caption{Charged particle track in the drift volume of the Triple-GEM detector.}
\label{fig: track_3gem}
\end{figure}

An ionizing particle crossing a portion of the active volume creates electrons and ions that move in opposite directions separated by the electric field.
The speed of the $\alpha$ particles and $^7$Li ions is much larger than the electron drift velocity, therefore the primary charge can be assumed to be released instantaneously along their path. The electron cloud moves rigidly at a constant speed due to the uniform electric field. Neglecting the small amount of diffusion, the electron cloud preserves its original track shape along the drift. The x and y view of one typical track as measured with the Triple-GEM setup is depicted in figure~\ref{fig: track_3gem}. Knowing the instant when the primary charge is released ($t_0$), one can extract the drift time and consequently the position along the drift of the ionizing event. In the present configuration this information is not accessible. Nevertheless, the electrons from the end of the track will always induce the signal before the electrons from the beginning of the track. In fact, since the converter is on the cathode, the beginning of the track has always the largest drift time. This concept forms the base of the Time Projection Chamber (TPC)~\cite{TPC}, which allows the identification of the beginning of the track even without knowing $t_0$. 

\begin{figure}
\centering
\begin{subfigure}[b]{.5\textwidth}
\centering
\includegraphics[width=\linewidth]{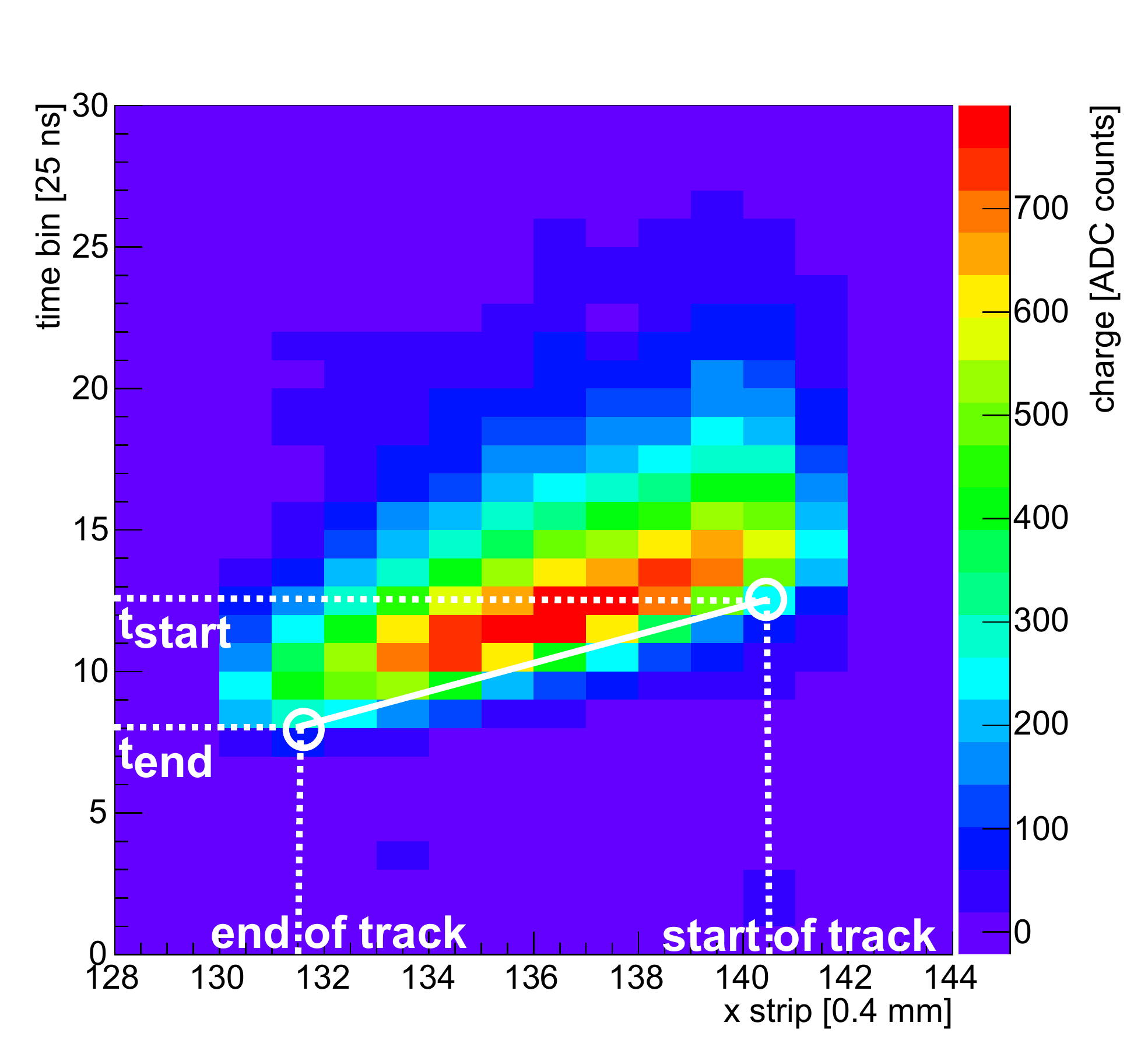}%
\caption{x view}
\label{fig: track_x_1gem}
\end{subfigure}%
\begin{subfigure}[b]{.5\textwidth}
\centering
\includegraphics[width=\linewidth]{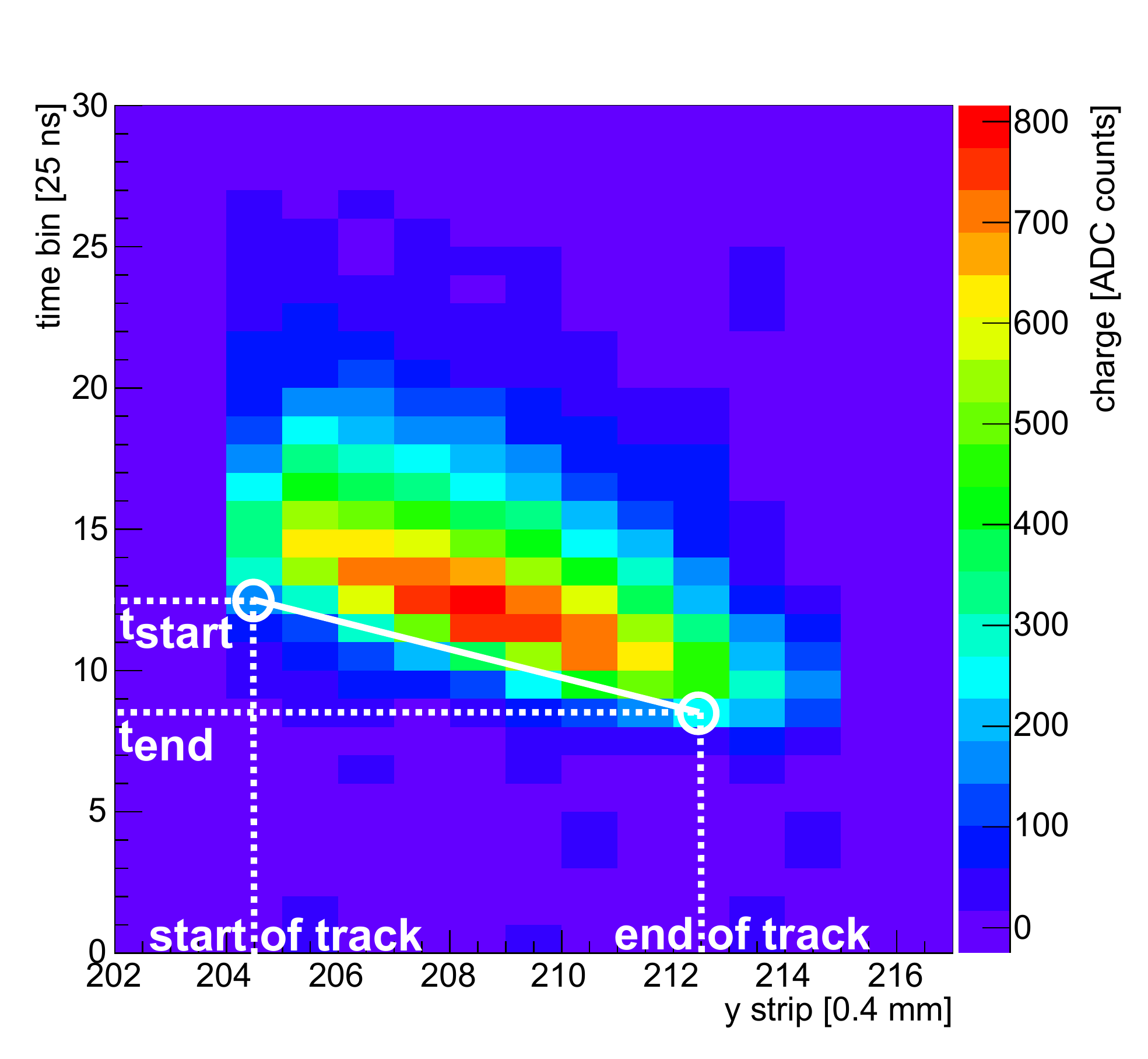}%
\caption{y view}
\label{fig: track_y_1gem}
\end{subfigure}
\caption{Charged particle track in the drift volume of the Single-GEM detector.}
\label{fig: track_1gem}
\end{figure}

The track in the Triple-GEM detector in figure~\ref{fig: track_3gem} and the track in the Single-GEM detector in figure~\ref{fig: track_1gem} have a comparable inclination. But one notices a larger time difference between the start and the end of the track in the Triple-GEM detector, due to the lower drift speed. Changing the drift speed via the drift field or the gas mixture in the detector can thus modulate the desired time difference.

\begin{figure}
\centering
\begin{subfigure}[b]{.47\textwidth}
\centering
\includegraphics[width=\linewidth]{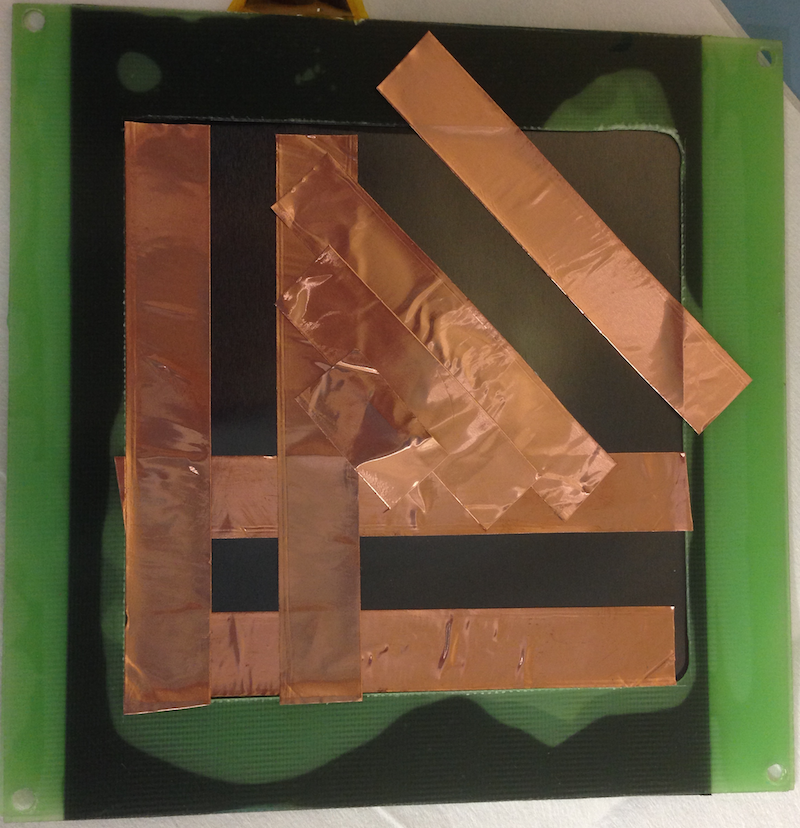}%
\caption{$^{10}$B$_4$C coated cathode}
\label{fig: cathode}
\end{subfigure}%
\begin{subfigure}[b]{.53\textwidth}
\centering
\includegraphics[width=\linewidth]{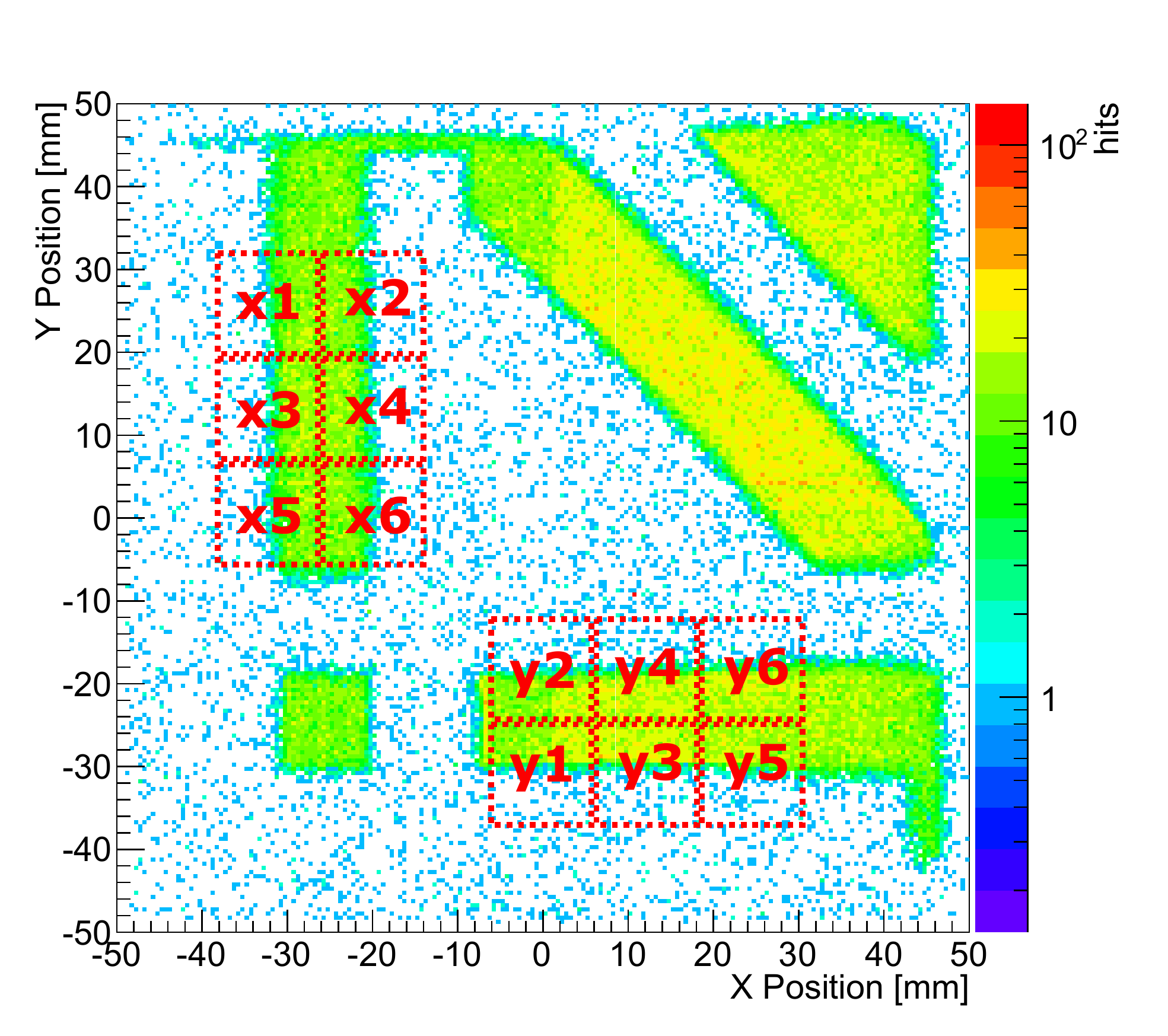}%
\caption{Hit distribution with analysed positions}
\label{fig: hit_distribution}
\end{subfigure}
\caption{$^{10}$B$_4$C coated cathode with copper tape and resulting hit distribution of the measurements with the Single-GEM.}
\label{fig: single_gem_hits}
\end{figure}

Figure~\ref{fig: cathode} shows the cathode with copper grid that was used for the Single-GEM measurement. The copper grid was added to stop the $\alpha$ particles and $^7$Li ions from escaping from the converter into the gas. This makes it possible to measure position resolution with a highly divergent source. As a result, sharp boundaries between the regions with and without recorded hits are visible in the plot of the hit distribution. To determine the improvement in position resolution due to the $\mu$TPC analysis, regions with a large step in the hit distribution as displayed in figure~\ref{fig: hit_distribution} were studied. Due to the substantial gamma flux of the $^{241}$AmBe source, a certain amount of gamma background was present during the measurements as indicated by the hits in areas covered with copper tape. To exclude the gamma background, which deposits a lot less charge than $\alpha$ particles and $^7$Li ions, an amplitude threshold was applied to the data. Subsequently a simple constant fraction discriminator was used to determine the time at which the signal in each strip reached 50 $\%$ of the maximum amplitude. The strip with the highest threshold crossing time was then chosen as the position of the start of the track.

\begin{figure}[htbp]
\centering
\includegraphics[width=0.95\textwidth]{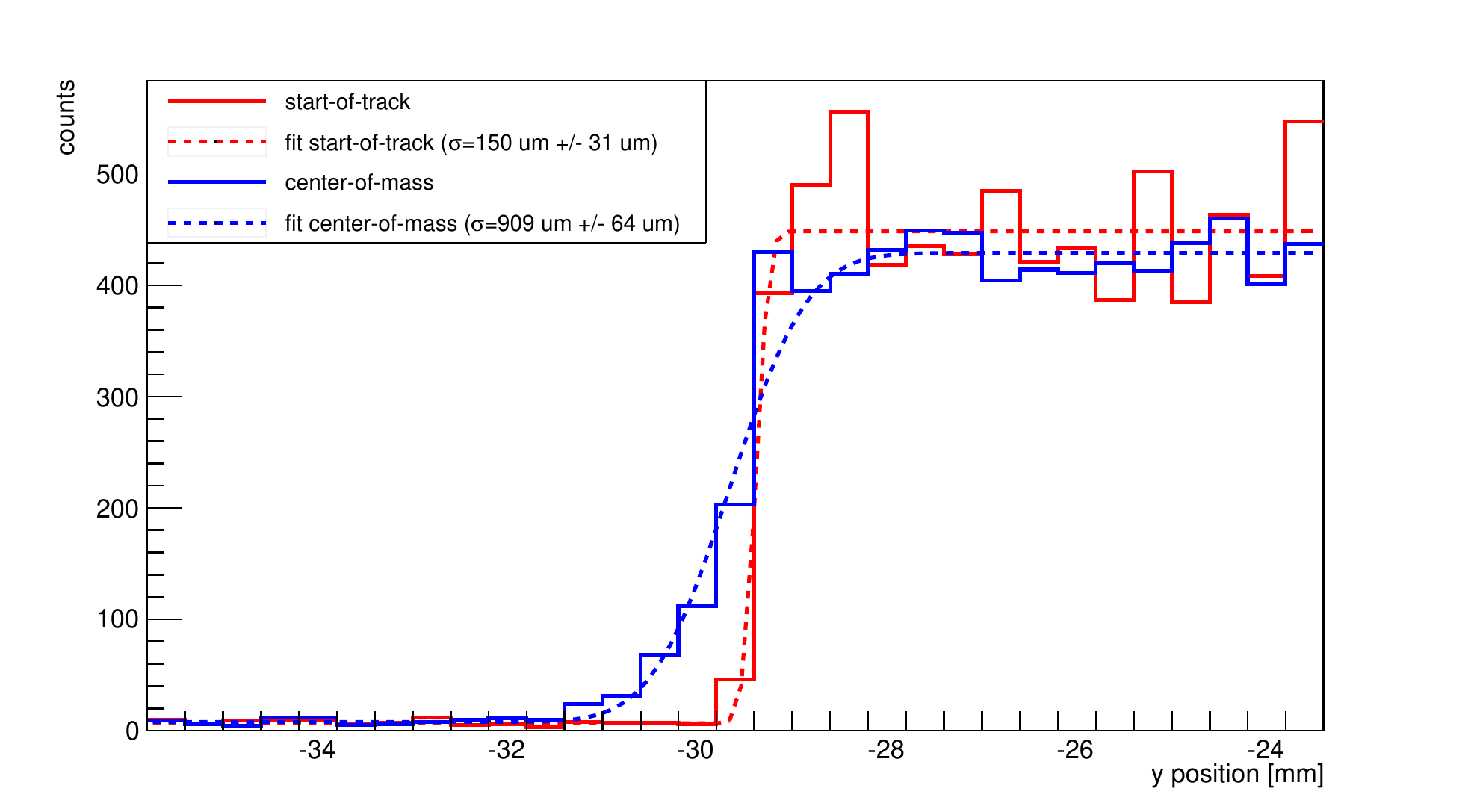}
\caption{Single-GEM position y5: Distribution of the reconstructed y coordinate using a center-of-mass-based technique (in blue) and the TPC analysis (in red). }
\centering
\label{fig: fit_1gem_y}
\end{figure}

Figure~\ref{fig: fit_1gem_y} shows as example the analysis of the position resolution in \emph{region y5} in figure~\ref{fig: hit_distribution}. The standard deviation $\sigma$ of the position resolution is extracted from the fit of the complementary error function to the data. The reconstructed position obtained with the $\mu$TPC concept is shown in red, the centre-of-mass approach in blue. The $\sigma$ of the position resolution amounts to 150~$\mu$m for the $\mu$TPC concept and 909~$\mu$m for the centre-of-mass approach. The results of all analysed regions can be found in table~\ref{table: results}. The $\sigma$ of the position resolution is everywhere better than 270~$\mu$m for the Single-GEM measurements and the $\mu$TPC technique. The exact value depends on the flatness and straightness of the applied tape. In regions with straight and flat tape, the  $\sigma$ is 200~$\mu$m or smaller. 200~$\mu$m  or half the size of the strip pitch seems to be a limitation for this fit method. As expected, if the $\sigma$ gets too small with respect to the readout pitch, the fit cannot converge as the sharp drop-off occurs in < 1 strip.

The strips of the cartesian x/y readout have a 400 $\mu$m pitch with 80 $\mu$m strip width in y, and 340 $\mu$m strip width in x. Since the x strips are underneath the y strips and electrically separated from them by 50~$\mu$m of Kapton, the different strip sizes have been chosen to have a more equal charge sharing between y and x. But as the data analysis shows, the y strips on the top get about 60$\%$ of the charge, whereas the x strips get around 40$\%$. Due to the larger amount of charge,  the results for the y coordinate are in general better than for the x coordinate. This is valid for both the $\mu$TPC technique and the centre-of-mass approach. During the Triple-GEM measurements half of the detector area was covered by a 25~$\mu$m thick aluminium foil with a solid layer of $^{10}$B$_4$C. The edge of the aluminium foil was not very smooth and not well aligned with the strips of the readout board. The $\sigma$ of the obtained position resolution amounted therefore to 311~$\mu$m.

\FloatBarrier
\newpage

\section{Conclusions}
\label{sec:conclusions}
The $\mu$TPC  concept is shown for the first time with thermal neutrons using $^{10}$B$_4$C as neutron converter in a Single-GEM and Triple-GEM detector configuration. By using a simple laboratory setup with a radioactive neutron source and copper tape on the cathode to mask the neutron converter and give a sharp edge, it is possible to extract the position resolution. These proof-of-principle measurements demonstrate that the resulting position resolution is improved by a factor of ca 5 using a $\mu$TPC technique compared to a more traditional centre-of-mass based method. The obtained resolutions are with < 200~$\mu$m better than half of the readout strip pitch. The $\mu$TPC concept thus fulfils the required position resolution for NMX. It also presents a perspective for time-resolved neutron detectors with spatial resolutions far below 1~mm by applying this technique. Measurements under better conditions such as a higher flux of thermal neutrons and a lower gamma background are foreseen to further evaluate the potential of the $\mu$TPC technique. 
 
\begin{table}
\centering
\begin{threeparttable}
\begin{tabular}[htbp]{ l|r|r|r|r|}
\cline{2-5}
& \multicolumn{2}{ |c| }{start-of-track} & \multicolumn{2}{ |c| }{center-of-mass}  \\ 
\hline
\multicolumn{1}{ |c|}{Position} & $\sigma$ [$\mu$m] & +/- error [$\mu$m] & $\sigma$ [$\mu$m] & +/- error [$\mu$m]  \\ 
\hline
\multicolumn{1}{ |c|}{Triple-GEM} & 311 & 61 & 1161 & 111 \\ 
\multicolumn{1}{ |c|}{Single-GEM x1} & < 200\tnote{*} & - & 1350 & 67   \\ 
\multicolumn{1}{ |c|}{Single-GEM x2} & < 200\tnote{*} & -  & 1246 & 71   \\ 
\multicolumn{1}{ |c|}{Single-GEM x3} & 238 & 17 & 1113 & 69  \\ 
\multicolumn{1}{ |c|}{Single-GEM x4} & < 200\tnote{*} & - & 1213 & 62 \\ 
\multicolumn{1}{ |c|}{Single-GEM x5} & 268 & 24 & 1498 & 67  \\ 
\multicolumn{1}{ |c|}{Single-GEM x6} & < 200\tnote{*} & -  & 1263 & 66  \\ 
\multicolumn{1}{ |c|}{Single-GEM y1} & < 200\tnote{*} & -  & 593 & 59  \\ 
\multicolumn{1}{ |c|}{Single-GEM y2} & < 200\tnote{*} & -  & 942 & 55   \\ 
\multicolumn{1}{ |c|}{Single-GEM y3} & 146 & 26 & 587 & 49  \\ 
\multicolumn{1}{ |c|}{Single-GEM y4} & 116 & 69 & 786 & 56 \\ 
\multicolumn{1}{ |c|}{Single-GEM y5} & 150 & 31 & 909 & 64 \\ 
\multicolumn{1}{ |c|}{Single-GEM y6} & < 200\tnote{*} & -  & 841 & 59  \\ 
\hline
\end{tabular}
\caption{Fit results for the Triple-GEM and the 12 Single-GEM regions of interest depicted in figure~\protect\ref{fig: hit_distribution}.}
\label{table: results}
\begin{tablenotes}
\item[*] Limit of fit reached.
\end{tablenotes}
\end{threeparttable}
\end{table}

\FloatBarrier

\end{document}